\documentclass[prl,twocolumn,showpacs,preprintnumbers]{revtex4}
\usepackage{graphicx}
\usepackage{dcolumn}

\begin{document}

\title{On the heat current in the magnetic field: \\
Nernst-Ettingshausen effect above the superconducting transition}

\author{ A. Sergeev}
\email{asergeev@eng.buffalo.edu} \affiliation{ Research
Foundation, University at Buffalo, Buffalo, New York 14260}
\author{M. Yu. Reizer} \affiliation{5614 Naiche Rd.
Columbus, Ohio 43213}
\author{V. Mitin} \affiliation{Electrical Engineering
Department, University at Buffalo, Buffalo, New York 14260}


\begin{abstract} For maintaining gauge invariance in
a magnetic field, the heat current operator should include the
magnetic term. Taking this term into account, we revised
calculations of the Nernst-Ettingshausen effect above the
superconducting transition. We found that the fluctuations of the
modulus of the order parameter do not change the particle-hole
asymmetry (PHA) of the thermomagnetic effects. As in the normal
state,  the thermomagnetic effects in the fluctuation region are
proportional to the square of PHA and, therefore, small. Large
Nernst effect observed in the high-$T_c$ cuprates requires
vortex-like excitations due to the phase fluctuations, which are
beyond the Gaussian-fluctuation theory.
\end{abstract}


\maketitle The observation of large thermomagnetic effects in
cuprates turns out to be crucial for understanding high-$T_c$
superconductivity \cite{1,2,3}. In the Nernst effect, the electric
current is induced by crossed temperature gradient and magnetic
field, ${\bf j}^e = N[\nabla T \times {\bf H}]$. In an open
circuit, the Nernst voltage $E_y$ is given by
\begin{eqnarray}\label{EN}
  {E_y \over (-\nabla T)_x H_z} = {N_{xy}\over \sigma_{xx}} -
  {\eta_{xx} \sigma_{xy} \over \sigma_{xx}^2},
\end{eqnarray}
where $\eta_{xx}$ is the thermoelectric coefficient, $\sigma_{xx}$
and $\sigma_{xy}$ are the electrical and Hall conductivities.

In metals, the Nernst voltage is small, because it requires strong
particle-hole asymmetry (PHA) in the electron systems. PHA of the
thermoelectric coefficient $\eta_{xx}$ is given by the factor
$T/\epsilon_F$, where $\epsilon_F$ is the Fermi energy; PHA of the
Hall conductivity $\sigma_{xy}$ may be presented as $\Omega_H
\tau$, where $\Omega_H = eH/mc$ and $\tau$ is the electron
momentum relaxation time. Coefficient $N_{xy}$ contains both PHA
factors of $\eta_{xx}$ and $\sigma_{xy}$. Thus, in Eq. \ref{EN}
both terms are proportional to the square of PHA. Moreover, if
$\tau$ is independent on energy, they cancel each other.

The important result of Ref. \cite{1} is that the large measured
Nernst voltage originates from $N_{xy}$, i.e. the first term in
Eq. \ref{EN} strongly dominates over the second term in the
temperature range extending from below $T_c$ to a broad interval
above. It now seems clear that the region above $T_c$ is in fact
the vortex fluid phase. Various phenomenological models are
suggested to describe this phase and heat transfer by vortices
\cite{4,5,6,7}. It is also believed (see a review \cite{7}), that,
at least close to $T_c$, the Nernst voltage is described by the
microscopic Gaussian-fluctuation theory, which predicts the finite
$N_{xy}$ in the zero order in PHA \cite{8,9,10,11}. Below we will
show that this statement is wrong.  In the Gaussian model with
ordinary particle-hole excitations, it is not possible to change
the PHA of $N_{xy}$. To obtain the large Nernst effect due to
vortex-like excitations, it is necessary to go beyond fluctuations
of the modulus of the order parameter.

The Gaussian-fluctuation theory allows one to calculate the
Ettingshausen coefficient $\Upsilon$, which describes the heat
flux induced by crossed electric and magnetic fields, ${\bf j}^h =
-\Upsilon [{\bf E} \times {\bf H}]$. According to the Onsager
relation,
\begin{eqnarray}\label{Ons}
N=\Upsilon/T
\end{eqnarray}

Results of the previous theoretical works can be formulated by two
statements. First, in the infinite sample the coefficient
$\Upsilon$ near $T_c$ does not require PHA,
\begin{eqnarray}\label{Us}
\delta \Upsilon_{inf} =  -{e^2 T \over 4\pi c} \times \cases{
    2 \alpha/\eta \ \ \ \propto \ (T-T_c)^{-1}
    & for 2D, \cr
    \sqrt{\alpha/\eta} \ \ \propto \ (T-T_c)^{-1/2}
    & for 3D, \cr}
 \end{eqnarray}
where $\eta=(T-T_c)/T_c$ and $\alpha = (T-T_c)[\xi (T)]^2 $, $\xi
(T)$ is the coherence length. For the first time, Eq. \ref{Us} has
been obtained in Ref. \cite{8} using the time-dependent
Ginzburg-Landau (TDGL) theory. Later it has been reproduced in
Ref. \cite{10}. Finally, Eq. \ref{Us} has been confirmed by the
diagrammatic calculations in the frame of the Aslamazov-Larkin
(AL) approximation \cite{11}.

Second, according to Refs. \cite{10,11,12} in the finite sample
the Eq. \ref{Us} should be corrected due to surface currents
related to thermal magnetization currents,
\begin{eqnarray}\label{mag}
\delta \Upsilon &=& \delta \Upsilon_{inf} - c\mu, \\ \mu &=& -
{e^2 T \over 6 \pi c^2} \times \cases{
   2 \alpha/\eta
    & for 2D, \cr
    \sqrt{\alpha/\eta}
    & for 3D. \cr}
\end{eqnarray}
where the $\mu$ is the fluctuation magnetic susceptibility
\cite{13}. This correction would decrease the $\Upsilon$ to one
third of its value given by Eq. \ref{Us}.

In the current paper we develop consistent microscopic description
of the heat current in the magnetic field. Calculating $\delta
\Upsilon$ by the Kubo method and $\delta N$ by the transport
equation, we show that both statements expressed by Eqs. \ref{Us}
and \ref{mag} are wrong. Our conclusion is that in the frame of
Gaussian-fluctuation theory the coefficients $\delta \Upsilon=
\delta N=0$ in the the zero order in PHA.

We start our calculations with discussion of the energy and heat
current operators. The energy current of interacting electrons may
be obtained from the equation of motion for the electron field
operators or from the energy-momentum tensor \cite{14}. For
electrons interacting in the Cooper channel, the energy and heat
current operators have been derived in our work \cite{15}. In the
external fields ${\bf H}= i [ {\bf k}_H \times {\bf A}_H ]$ and
${\bf E}= -i {\bf k}_E \phi $, the electron energy has a
gauge-invariant form
\begin{eqnarray}\label{HH}
 \tilde{E}=({\bf p}+e{\bf A}_H/c)^2/2m +e\phi.
\end{eqnarray}
 The energy current operator is easily generalized to include the
external fields,
\begin{eqnarray}\label{Jh}
{\hat {\bf J}}^\epsilon = \sum_{\bf p} {\bf v} \xi_p  a_{\bf p}^+
a_{\bf p} + \sum_{\bf p} {e{\bf v}\over c} ({\bf v}{\bf A}_H)
a_{\bf p}^+a_{\bf p} + \sum_{\bf p} {\bf v} e\phi  a_{\bf p}^+
a_{\bf p} \nonumber\\ + \sum_{\bf p} {\bf v} \mu \ a_{\bf p}^+
a_{\bf p} - {\lambda /2} \sum_{{\bf p},{\bf p}',{\bf p}''} ({\bf
v}+{\bf v}') \ a_{{\bf p}+{\bf p}'-{\bf p}''}^+ a_{{\bf p}''}^+
a_{{\bf p}'} a_{\bf p} \nonumber \\ + \sum_{{\bf p},{\bf p}',{\bf
R}_i} ({\bf v}+{\bf v}') U_{imp}({\bf R}_i) \ \exp [i({\bf p}+{\bf
p}') R_i] \ a_{\bf p}^+ a_{{\bf p}'} \ , \ \ \
\end{eqnarray}
where $a_{\bf p}^+$ and $a_{\bf p}$ are the electron creation and
annihilation operators, $\xi_p = p^2/2m -\mu $ ($\mu$ is the
chemical potential), $U_{imp}$ is the impurity potential. The
first and the forth terms in the Eq. \ref{Jh} describes the energy
flux of noninteracting electrons without fields. The second and
third terms are due to electron interaction with magnetic and
electric fields. Two last terms are many-body corrections due to
the electron-electron and electron-impurity interations. According
to Ref. \cite{15}, these two terms generate diagram blocks
(AL-blocks) proportional to PHA, but contributions of all blocks
with PHA cancel each other. Anyway, these many-body corrections
can be neglected, if $\Upsilon$ is calculated in the zero order in
PHA.

The thermal energy can be defined as the electron energy counted
from the electro-chemical potential $e\phi+\mu$. Then the heat
current may be presented in terms of the energy and electric
currents as ${\bf J}^h = {\bf J}^\epsilon - (\mu / e+ \phi) {\bf
J}_e$. Obviously, the third and the forth terms in Eq. \ref{Jh} do
not contribute to the heat current. Thus, calculating the heat
current we should take into account only two heat current vertices
$\gamma^h_1$ and $\gamma^h_2$ corresponding to the first (kinetic)
and second (magnetic) terms in Eq. \ref{Jh}.

In the linear response method, the thermoelectric coefficient is
given by the correlator of the heat and charge currents \cite{12}.
Calculating response to ${\bf E} \times {\bf H}$, it is convenient
to use the gauge conditions ${\bf k}_H \cdot {\bf A}_H= {\bf k}_E
\cdot {\bf A}_H= 0$, in this case ${\bf E} \times {\bf H}= -i{\bf
A}_H({\bf k}_H\cdot{\bf E})$.

Two leading terms in the heat current operator generate two
diagrams, which describe $\Upsilon$ in the Aslamazov-Larkin (AL)
approximation. The left blocks ${\bf B}^h_1$ and ${\bf B}^h_2$ in
Figs. 1.a and 1.b  are blocks of electron Green functions
connected with heat current vertices $\gamma^h_1$ (kinetic) and
$\gamma^h_2$ (magnetic). General for both diagrams, the right
block ${\bf B}^e$ includes the electric current vertex $\gamma^e=
e{\bf v}\cdot{\bf e}_E $, ${\bf e}_E = {\bf E}/E $. Block ${\bf
B}^H$ includes the magnetic vertex $\gamma^H = (e/c){\bf
v}\cdot{\bf A}_H$. The wavy lines correspond to the fluctuation
propagator, which is given by \cite{13,15}
\begin{equation}\label{L1}
L^{R,A}(q,\omega) =\bigl(\lambda^{-1}-P^{R,A}(q,\omega)
\bigr)^{-1},
\end{equation}
where  $\lambda$ is a constant of the interaction in the Cooper
channel and $P(q,\omega)$ is the polarization operator,
\begin{eqnarray}\label{P2}
P^{R,A}(q,\omega)= -{\nu \over 2} \biggl (\ln{{2\ C_\gamma
\omega_D \over \pi T}} -\alpha q^2 \pm {i\pi \omega  \over 8T}
+\gamma \omega \biggr),
\end{eqnarray}
$\nu$ is the electron density of states, $\omega_D$ is the Debye
frequency, and $C_\gamma$ is the Euler constant. The last term in
Eq. \ref{P2} is proportional to PHA and can be ignored in zero
order in PHA calculations.

For an arbitrary electron momentum relaxation time $\tau$, the AL
blocks ${\bf B}^{e,h,H}$ of electron Green functions $G^{R(A)}$
with vertices $ \gamma^e $, $ \gamma^H $, and  $ \gamma^h_i $ are
given by \cite{13,15}
\begin{eqnarray}\label{B}
{\bf B}^{e,h,H}_i = {\rm Im}  \int {d {\bf p} \ d \epsilon \over
(2 \pi)^4} \ {\bf v} \ \gamma^{e,h,H}_i \ S(\epsilon) \ {(G_p^A)^2
G^R_{q-p} \over (1- \zeta)^2}, \\  S(\epsilon)= -\tanh
{\epsilon\over 2T}, \ \zeta =  {1\over  \pi \nu \tau} \int {d{\bf
p} \over (2\pi )^3} \ G^A_p G^{R}_{q-p},
\\ G^R_p = [G^A_p]^* = (\epsilon-\xi_p+i/2\tau)^{-1}.
\end{eqnarray}
The block ${\bf B}^e$ with the electric current vertex, $\gamma^e=
e{\bf v}\cdot{\bf e}_E $, may be presented as \cite{13,15}
\begin{eqnarray}\label{Be}
{\bf B}^e({\bf q}) \ = \ 2e \ \nabla_{\bf q} P^R({\bf q},0)\cdot
{\bf e}_E= \ 2e\nu\alpha \ {\bf q}\cdot {\bf e}_E.
\end{eqnarray}
The block ${\bf B}^H$ with the vertex  $\gamma^H = (e/c){\bf
v}\cdot{\bf A}_H$ is
\begin{eqnarray}\label{BA}
 {\bf B}^A({\bf q}) \ = \ 2(e/c)\nu\alpha \
{\bf q}\cdot {\bf A}_H.
\end{eqnarray}
The block ${\bf B}^h_1$ with the kinetic heat current vertex,
$\gamma^h_1=
 \xi {\bf v}\cdot {\bf e}_{j^h} $ (${\bf e}_{j^h} =
 {\bf j}^h/j^h  \ \| {\bf A}$),
is given by \cite{15} (see also \cite{10})
\begin{eqnarray}\label{Bh1}
{\bf B}^h_1({\bf q},\omega) \ = \ \omega \ \nabla_{\bf q} P^R({\bf
q},0)\cdot {\bf e}_{j^h}= \omega \nu\alpha \ {\bf q}\cdot {\bf
e}_{j^h}.
\end{eqnarray}
Next,  we calculate the block ${\bf B}^h_2$ with the magnetic heat
current vertex $\gamma^h_2= ({\bf v} \cdot {\bf A}_H) ({\bf
v}\cdot {\bf e}_{j^h})$. The integral over angles of the electron
momentum involves only the vertex $\gamma^h_2$, because the heat
current is in the direction of ${\bf A}_H$. To obtain an imaginary
part in Eq. \ref{B}, the integral
\begin{eqnarray}
 \int d\xi \ (G_p^A)^2 G^R_{q-p}= {2 \pi i \over (2\epsilon-\omega
 -{\bf q}\cdot {\bf v}-i/\tau)^2},
\end{eqnarray}
should be expanded in $\omega$ (in calculations of ${\bf B}_e$ it
is expanded in ${\bf q}\cdot {\bf v}$). Finally, we get
\begin{eqnarray}\label{Bh2}
{\bf B}^h_2({\bf q}, \omega)  =  2A_H \ {\omega \over {\bf q}} \
\nabla_{\bf q} P^R({\bf q},0)=2 (e/c) \omega   \nu \alpha A_H.
\end{eqnarray}

The first AL diagram (Fig. 1.a) for the Ettingshausen coefficient
is based on the blocks ${\bf B}^h_1$ and ${\bf B}^e$, the vector
potential ${\bf A}$ enters the fluctuation propagator. This
diagram was calculated in Ref. \cite {11} and its contribution to
$\Upsilon_{inr}$ is given by Eq. \ref{Us}. The same result has
been obtained in the TDGL formalism in Refs. \cite{8,10}.

The second AL diagram (Fig. 1.b) is based on the blocks ${\bf
B}^h_2$ and ${\bf B}^e$. The analytical expression corresponding
to this diagram is
\begin{eqnarray}\label{D2}
\Upsilon_{inr}^{(2)}H = \Im  \int {d{\bf q} \over (2\pi ) ^n}
{d\omega \over 2\pi} \ {{\bf B}^h_2  {\bf B}^e\over 2 \Omega}
(L^C_+L^A_- +L^R_+ L^C_-), \
\end{eqnarray}
where $L^C=\coth(\omega/2T)(L^R-L^A)$, $L_\pm$ is used for $L({\bf
q} \pm {\bf k}/2, \omega \pm \Omega/2)$, and $n$ is the system
dimensionality with respect to the coherence length $\xi (T)$.
Expanding the integrant to the linear order in $\Omega$ and {\bf
k} and calculating the integrals over $\omega$ and ${\bf q}$, we
find that the contribution of the second diagram,
$\Upsilon_{inr}^{(2)}$ cancels completely the contribution of the
first one (Eq. \ref{Us}). The analogous cancellation of two
diagrams with the heat current vertices $\gamma^h_1$ and
$\gamma^h_2$ takes place in the case of noninteracting electrons
(see Appendix in Ref. \cite{16}).

Thus, without PHA the Ettingshausen effect is absent,
$\Upsilon=0$. To get nonzero result, we should expand the
fluctuation propagator (Eq. \ref{L1}) up to the second order in
PHA (in the first order in PHA both diagrams give zero results),
i.e. we expand the polarization operator (Eq. \ref{P2}) to the
second order in $\gamma \omega$. In the second order in PHA, the
thermomagnetic coefficients are
\begin{eqnarray}\label{Ett}
{\delta \Upsilon \over T} =  \delta N = - {5 e^2  \over 4\pi
c}\biggl({8T \gamma\over \pi }\biggr)^2  \cases{
    2 \alpha/\eta
    & for 2D, \cr
    \sqrt{\alpha/\eta}
    & for 3D; \cr}
 \end{eqnarray}
where ${\displaystyle \gamma = {1 \over 2 \epsilon_F} {\partial
\ln \nu \over
\partial \ln \epsilon_F}  \ \ln {2 C_\gamma \omega_D \over \pi
T_c}} \ $ \cite{17}. Thus, the Ettingshausen and Nernst
coefficients in the fluctuation region are proportional to
$(T/\epsilon_F)^2$. Taking into account that in this region the
thermoelectric coefficient $\eta_{xx}$ and the Hall conductivity
$\sigma_{xy}$ are proportional to $(T/\epsilon_F)$, we see that in
Eq. \ref{EN} for the Nernst voltage both terms are of the same
order, ~$(T/\epsilon_F)^2$.

The above results obtained for the infinite sample (or in the
interior of the sample) do not show any contribution of the
magnetization currents and, therefore, in the finite sample no
additional corrections are required. Because the role of
magnetization currents is widely misinterpreted in literature,
below we briefly analyze the energy current induced by the
magnetization current and show that this term is erroneously
attributed to the heat flux (Eq. \ref{mag}). In the potential
$\phi ({\bf r}) $, the electric magnetization current ${\bf
j}^e_{mag}$ transfers the energy flux ${\bf
j}^\epsilon_{mag}=\phi{\bf j}^e_{mag}$ (Eq. 37 in \cite{18}).
Taking into account that ${\bf j}^e_{mag} = c \mu k^2 {\bf A}_H $,
we get
\begin{eqnarray}\label{magcur}
{\bf j}^\epsilon_{mag}= c\mu[{\bf H}\times{\bf E}].
 \end{eqnarray}
In the linear response method, this term is given by the diagram
with the energy current vertex $\gamma^\phi= e \phi$ {\bf v} (the
third term in the energy current operator, Eq. \ref{Jh}) and the
magnetic vertex $\gamma^A = (e/c)({\bf A}_H \cdot{\bf v})$.
Without $\phi$ this diagram gives the ${\bf j}^e_{mag}$, and,
therefore, we get ${\bf j}^\epsilon_{mag}=\phi{\bf j}^e_{mag}$. As
we discussed, the vertex $\gamma^\phi= e{\bf v} \phi$ does not
contribute to the heat current, because the heat energy is counted
from the electro-chemical potential.

While the magnetization currents do not contribute to the heat
transfer, they play an important role in the charge transfer. In
the interior of the sample the electric current consists of the
transport and magnetization components, ${\bf j}^e_{inr} = {\bf
j}^e_{tr}+{\bf j}^e_{mag}$. According to Ref. \cite{18},
\begin{eqnarray}\label{jem}
{\bf j}^e_{mag} = c \ {\partial \mu \over \partial T} \ (\nabla T
\times {\bf H} ).
\end{eqnarray}
The magnetization currents are divergence-free. Therefore, the
total magnetization current through the sample cross-section must
be zero, i.e. the bulk magnetization currents are canceled by the
surface currents. Therefore, in the finite sample, the measured
Nernst coefficient $N$ is determined by the transport currents,
$N={\bf j}^e_{tr}/[\nabla T \times {\bf H}]$ \cite{18}. Calculated
from the theory, the Nernst coefficient in the infinite sample may
be associated with the current in the interior of the finite
sample, $N_{inf}={\bf j}^e_{inr}/[\nabla T \times {\bf H}]$. Using
Eq. \ref{jem}, we get
\begin{eqnarray}\label{Ons1}
\delta N = {{\bf j}^e_{inr} -{\bf j}^e_{mag} \over [\nabla T
\times {\bf H}]} = \delta N_{inf} - c {\partial \mu \over
\partial T}.
\end{eqnarray}

The coefficient $\delta N_{inf}$ may be calculated by the quantum
transport equation, where $\nabla T$ can be easy incorporated
\cite{14,15,16,19,20} (as a nonmechanical perturbation, $\nabla T$
cannot be incorporated into the Kubo formalism). In the AL
approximation, calculation of the Nernst coefficient is analogous
to calculation of the Hall effect \cite{19} with the only
difference that the derivatives of the fluctuation propagator
$L^{R,A}(q,\omega)$ with respect to $\omega$ in the Hall effect
should be replaced by the temperature derivative in calculations
of $\delta N_{inf}$. Taking into account temperature dependencies
of all coefficients in the fluctuation propagator (Egs. \ref{L1}
and \ref{P2}), for the 2D-superconductor we get
\begin{eqnarray}\label{Ninf}
\delta N_{inf} = \delta N+{e^2 \over 3 \pi c} \biggl( {\alpha
\over \eta^2} -{\alpha \over \eta}- {\alpha \over \eta} {\partial
\alpha \over \partial T}\biggl).
\end{eqnarray}
Thus, in the infinite sample or in the interior of the finite
sample, the Nernst  coefficient consists of a large term (the
second term in Eq. \ref{Ninf}), which is nonzero to zero order in
PHA. However, in the finite sample this term is canceled by the
contribution of the magnetization currents,  $c \partial \mu /
\partial T $. According to Eq. \ref{Ons1}, the rest is equal to
$\delta N$, which satisfies to the Onsager relation $\delta N=
\delta \Upsilon/T$.

Thus, in the finite sample both Nerst and Ettingshausen
coefficients in the fluctuation region (Eq. \ref{Ett}) as well as
the coefficients in the normal state ($N_0$ and $\Upsilon_0$) are
proportional to the square of PHA. Therefore, the relative
corrections do not consist of any PHA factors. For example, for 2D
superconductors we obtain
\begin{eqnarray}\label{rel}
{\delta N \over N_0} \sim {\delta \Upsilon \over \Upsilon_0} \sim
{1\over \eta} \cdot {1\over \epsilon_F \ {\rm max} \{ \tau, 1/T
\}}.
\end{eqnarray}
These corrections are of the same order as the relative
corrections to the conductivity, themoelectric and Hall
coefficients. Our consideration was limited by the one-band model.
However, Eq. \ref{rel} is also valid for the two-band model, where
the electron and hole contributions are additive and
thermomagnetic effects are large. In other words, our key
statement is that the interelectron interaction cannot change PHA
requirements for the thermomagnetic coefficients. The previous
works gave $\delta N/ N_0$ with an additional large factor of
$(\epsilon_F / T)^2$. However, none of experiments has shown such
huge effect in ordinary superconductors. While claiming opposite,
the very recent work on SiN \cite{21} actually supports our
results in the form of Eq. \ref{rel}, if the authors would have
normalized $\delta N$ obtained in the fluctuation region by $N_0$
taken from measurements far from the transition, say at $T \sim
2T_c$. In fact, they evaluated $N_0$ by the term $\eta_{xx}
\sigma_{xy} / \sigma_{xx}$, which was found to be many orders of
magnitude smaller than the measured $N$ even far above $T_c$.

Theoretical problems of thermomagnetic transport rests on
identification of the correct form of the heat current operator,
which takes into account the interaction corrections and
corrections due to external fields. As we have seen above, the
transfer of the potential energy $e \phi ({\bf r})$ by the
magnetization currents contribute to the energy current, but not
to the heat current, because the electron thermal energy is
counted from $\mu + e\phi$. Thus, the magnetization currents do
not affect the Ettingshausen effect. On the other hand, the
magnetic field gives an important contribution to the heat current
operator. Taking into account the magnetic term - viz., $
(e/c)({\bf v}\cdot {\bf A}_H) {\bf v}$ (Eq. \ref{Jh}), - we have
shown that the corresponding correction to the  Ettingshausen
coefficient cancels completely the result (Eq. \ref{Us}) obtained
with the kinetic term in the heat current operator. Thus, previous
calculations erroneously attributed the electric term in the
energy current to the heat current \cite{10,11,12} and overlooked
the magnetic term \cite{8,9,10,11}.

In summary, we have shown that any calculations of many-body
effects in thermomagnetic phenomena should include the magnetic
term in the heat current operator (the second term in Eq.
\ref{Jh}). Magnetization currents in the electric field contribute
to the charge and energy transfer, but not to the heat current.
Only in this way, one can obtained the Nernst and Ettingshausen
coefficients that satisfy to the Onsager relation (see Eqs.
\ref{Ons1} and \ref{Ninf}). The Gaussian fluctuations, i.e. the
fluctuations of the modulus of the order parameter, do not enhance
the thermomagnetic effects, because these fluctuations do not
change character of elementary excitations (particle-hole) and,
therefore, do not change the particle-hole asymmetry of the
thermomagnetic coefficients. The large Nernst effect observed in
high-$T_c$ cupprates requires vortex-like excitations related to
the phase fluctuations.

We would like to acknowledge useful discussions with I. Aleiner,
A. Larkin, D. Livanov, A. Varlamov, and I. Ussishkin.


\begin{thebibliography}{99}
\bibitem{1} Z.A. Xu, N.P. Ong, Y. Wang et al., Nature {\bf 406},
486 (2000).
\bibitem{2}
Y. Wang, S. Ono, Y. Onose et al.,  Science {\bf 299}, 86 (2003).
\bibitem{3}
Y. Wang, L. Li, and N.P. Ong, Phys. Rev. B. {\bf 73}, 024510
(2006).
\bibitem{4}
V.J. Emery and S.A. Kivelson, Nature (London) {\bf 374}, 434
(1995).
\bibitem{5}
O. Vafek and Z. Tesanovic, Phys. Rev. Lett. {\bf 91}, 237001
(2003).
\bibitem{6}
P.W. Anderson, cond-mat/0603726
\bibitem{7}
P.A. Lee, N. Nagaosa, X.G. Wen, Rev. Mod. Phys. {\bf 78}, 17
(2006).
\bibitem{8} S. Ullah and A.T. Dorsey, Phys. Rev. Lett. {\bf 65},
2066 (1990); Phys. Rev. B {\bf 44} 262 (1991).
\bibitem{9} A. A. Varlamov and D. V. Livanov,
Sov. Phys. JETP 72, 1016 (1991).
\bibitem{10}
I. Ussishkin, S.L. Sondhi, and D.A. Huse, Phys. Rev. Lett. {\bf
89}, 287001 (2002);
\bibitem{11}
I. Ussishkin, Phys. Rev. B {\bf 68}, 024517 (2003).
\bibitem{12}
S. Mukerjee and D.A. Huse, Phys. Rev. B {\bf 70}, 014506 (2004).
\bibitem{13}
 A. Larkin and A. Varlamov, {\it Theory of Fluctuations in
Superconductors}, Oxford Univ Press (2005).
\bibitem{14}
M. Reizer, A. Sergeev, J.W. Wilkins, and D. Livanov, Annals of
Phys. {\bf 257}, 44 (1997).
\bibitem{15}
M.Yu. Reizer and A.V. Sergeev, Phys. Rev. B {\bf 50}, 9344 (1994).
\bibitem{16}
M. Reizer and A. Sergeev, Phys. Rev. B. {\bf 61}, 7340 (2000).
\bibitem{17}
H. Fukuyama, H. Ebisawa, and T. Tsuzuki, Prog. Theor. Phys. {\bf
46}, 1028 (1971).
\bibitem{18} N.R. Cooper, B.I. Halperin, and I.M. Ruzin,
Phys. Rev. B {\bf 55}, 02344 (1997).
\bibitem{19}
A. Sergeev, M.Yu. Reizer, and V. Mitin, Phys. Rev. B. {\bf 66},
104504 (2002).
\bibitem{20}
G. Catelani and I. L. Aleiner JETP {\bf 100}, 331 (2005).
\bibitem{21}
A. Pourret, H. Aubin, J. Lesueur et al., Nature Phys. {\bf 2}, 683
(2006)

\end{thebibliography}
\end{document}